\documentclass[smallextended]{svjour3}
\usepackage{amssymb}
\usepackage{graphicx}

\RequirePackage{fix-cm}
\smartqed

\begin{document}

\title{Statistical signature of vortex filaments: dog or tail? Talk given at QFS 2016.\thanks{%
This work was supported by grant RFBR 15-02-05366} }
\author{Sergey K. Nemirovskii }
\date{Received: date / Accepted: date}
\maketitle

\begin{abstract}
The title of the paper coincides with the title of a paragraph in the famous
book by U. Frisch (1995)on classical turbulence. In this paragraph the author
discussed the role of statistical dynamics of vortex filaments in the theory
of turbulence and put the above question. In other words, whether the main
properties of turbulence (cascade, scaling laws) are the sequence of the
vortex line dynamics or the latter have only marginal signature. Quantum fluids, where the vortex filaments are the real objects, give an excellent opportunity to explore the role of discrete vortices in turbulent phenomena. The aim of this paper is to discuss which elements of vortex dynamics would lead to the
main ingredients of the theory of turbulence. We discuss how the nonlinear dynamics of vortex filaments can result in an exchange of energy between different scales, the formation of the Kolmogorov-type energy spectra and the decay of turbulence.\\
\end{abstract}



\titlerunning{Statistical signature} 


\institute{F. Author \at
                            Institute of Thermophysics, Lavrentyev ave, 1, 630090, Novosibirsk, Russia,\\
Novosibirsk State University, Novosibirsk Russia\\
                                          \email{nemir@itp.nsc.ru}            }


\section{Introduction.}

The idea that classical turbulence can be modeled by a set of slim vortex
tubes (or vortex sheets) has been discussed for quite a long time (See \cite%
{Frisch1995}). In classical fluids, the concept of thin vortex tubes is a
rather fruitful mathematical model. Quantum fluids, where the vortex
filaments are real objects, give an excellent opportunity for the study of
the question, whether the dynamics of a set of vortex lines is able to
reproduce the properties of real hydrodynamic turbulence. The main goal of
this paper is to discuss which elements of vortex dynamics would lead to
main ingredients of theory of turbulence.

In the paper we restrict ourselves to the most manifest features of the
turbulent flow, such as the exchange of energy between the different scales
and the Kolmogorov-type energy spectra We discuss the exchange of energy due
to recombinations (splitting and fusion) of the vortex loops and due to the
nonlinear stochastic deformation of initially smooth ring. We also describe
the attempts to obtain the Kolmogorov type spectra from the configurations
of vortex lines, which appeared as a result of the tangle evolution. In
particular, we investigate the reconnecting lines and inhomogeneous vortex
bundles.We also consider the problem of decay of turbulence and its relation
to the such phenomena as the KW cascade and the emission of small loops. And
finally we intend to summarize inquiring, whether outlined issues are
relevant to classical turbulence, or that are other phenomena.

\section{Exchange of energy between different scales. Vortex loops kinetics}

\textbf{Recombination of Gaussian loops.} This model assumes that the vortex
tangle consists of a set of many closed vortex loops, which undergo an
enormous number of reconnections and self-reconnections, reaching (for
typical experiments) values of order of several millions collisions per
second (per cm$^{3}$). Thus, in the full statement of the problem we have to
deal with a set of objects (vortex loops), which do not have a fixed number
of elements, they can be born and die. One of the possible treatments of
this problem is to impose the vortex loops to have the random walking
structure. The idea that the random one-dimensional topological defects have
a random walking structure is widespread (see, e.g., books by Kleinert \cite%
{Kleinert1990}). For vortex loops in superfluid helium this idea is realized
in form of the so-called generalized Wiener distribution, which takes into
account the anisotropy, polarization and the finite curvature. It has been
developed in \cite{Nemirovskii1998}. That approach allows to study kinetics
of vortex loops. The corresponding study was performed in works \cite%
{Nemirovskii2006,Nemirovskii2008,Nemirovskii2013}. To demonstrate the
existence of energy cascade in space of scales it was chosen the usual
variant of the Wiener distribution with an\ elementary step $\xi _{0}$ of
the order of intervortex distance $\xi _{0}\sim \delta =$ $\mathcal{L}%
^{-1/2} $, where $\mathcal{L}$ is the vortex line density (VLD).

\begin{figure}[tbp]
\includegraphics[width=8cm]{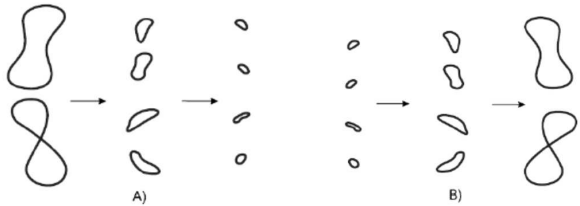}
\caption{(Color online) The cascade like breaking down and merging of
Gaussian vortex loop due to reconnection and self-reconnection processes. }
\label{recombination}
\end{figure}

Then, the only degree of freedom which exists is the length $l$ of the loop,
and all interactions reduce to collisions and recombinations (see Fig. \ref%
{recombination}). The corresponding problem can be studied on the basis of
the Boltzmann-type \textquotedblleft kinetic equation\textquotedblright\ and
has an exact, power-like solution for the distribution function of number
density of loops $n(l)$\ with length$\ l$, namely $n(l)=const\ast l^{-5/2}$.
This theory gives a series of predictions which can be associated with
quasi-classical behavior of quantum turbulence.

\textbf{The constant flux of the energy. }The found solution corresponds to
a nonequilibrium state, it describes a flux of length density $%
L(l,t)=ln(l,t) $, which is length, accumulated in loops of size $l$ in $l-$%
space. The term \textquotedblright flux\textquotedblright\ here means just
the redistribution of length among the loops due to reconnections.
Conservation of the vortex line density can be expressed in the form of a
continuity equation for the length density $L(l,t)$

\begin{equation}
\frac{\partial L(l,t)}{\partial t}+\frac{\partial P(l)}{\partial l}=0.
\label{continuity equation}
\end{equation}%
In the local induction approximation \ (LIA)\ the energy of a line is
proportional to its length, $E\propto \kappa ^{2}L$, therefore Eq.\ref%
{continuity equation} describes the constant flux of the vortex energy $%
P^{E}(l)=const$ in space of scales (which are just the loop sizes $l$). The
flux $P^{E}$ consists of two contributions. The first, positive one, is
related to merging of loops, and responsible for delivering the energy into
large scales. The second, negative contribution appears due to breaking-down
of loops (see Fig. \ref{recombination}) describes flux of energy to small
scales. Depending on the temperature either one oe the other can prevail, it
depends just on the temperature behavior of structure constants of the
vortex tangle. In particular, for $T=0$, \ energy flux $P^{E}<0$, the energy
is transferred into the region of small scales. This corresponds to the
direct cascade in classical turbulence.

\textbf{Effective Kinematic Viscosity. }We can rewrite the flux of energy
using VLD $\mathcal{L}$ (up to factor $(1/4\pi )\ln (1/\mathcal{L}%
^{1/2}a_{0})$, $a_{0}$- the vortex core size)%
\begin{equation}
P^{E}=C_{F}\kappa (\kappa \mathcal{L})^{2}.  \label{diss rate}
\end{equation}%
We named this constant $C_{F}$ in honor of Feynman, who was the first who
discussed the decay of the superfluid turbulence due to the cascade-like
breaking down of vortex loops. Expression of type (\ref{diss rate}) has the
widely accepted form (see. e.g., \cite{Vinen2010}). In case of the
counterflowing turbulence it appeared directly from the Vinen equation. In
the case of the quasi-classical turbulence, this form of (\ref{diss rate})
is a consequence of the fact that the dissipation rate is proportional to
the squared averaged vorticity $\left\vert \mathbf{\omega }\right\vert ^{2}$%
, and of the supposition that $\left\langle \left\vert \mathbf{\omega }%
\right\vert \right\rangle =\kappa \mathcal{L}$\emph{.}

\ We obtained relation (\ref{diss rate}) on a different ground. Referring
for details to works \cite{Nemirovskii2006,Nemirovskii2008,Nemirovskii2013}
we can estimate the coefficient in (\ref{diss rate}) as $C_{F}\ \approx
-0.25 $. This result agrees by the order with the value of the effective
kinematic viscosity $\nu ^{\prime }$, extracted from\textbf{\ }experiments
on decay of vortex tangle.

Resuming, it can be argued that the process of breaking down of vortex loops
can be associated with the main feature of turbulence -- the constant flux
of energy in space of scales, and the relation between this flux and VLD (%
\ref{diss rate}) agrees with experimental data.

\section{Exchange of energy between different scales. Stochastic deformation
of loops}

\label{NPP}

\begin{figure}[tbp]
\includegraphics[width=6cm]{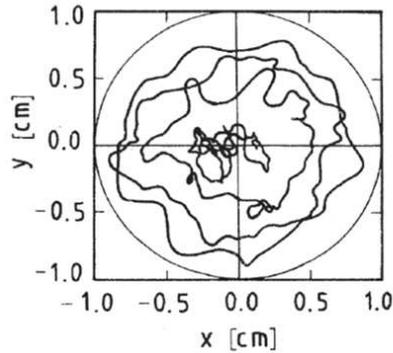}
\caption{(Color online) Evolution of vortex ring under influence of external
random force in local approach (Eq. \protect\ref{NPP eq}). As predicted, an
consequent arising of higher harmonics takes place leading eventually to an
entanglement of the initially smooth vortex loop. }
\label{loopdeformation}
\end{figure}

Another mechanism of exchange of the energy between scales, inherent for
turbulent phenomena, is related to nonlinear dynamics of a single vortex
filament (ignoring the reconnections.). Here I describe the solution of the
problem of chaotic distortions of a vortex loop \cite%
{Nemirovskii1991,Nemirovskii2001b}. In the LIA the chaotic motion of a
quantized vortex filament obeys the Langevin type equation
\begin{equation}
{\frac{{d~}\mathbf{s}{(\xi ,t)}}{{d~t}}}~=~\beta \mathbf{s}^{\prime }\times
\mathbf{s}^{\prime \prime }+\mathbf{\eta }(\xi ,t)+\mathbf{\zeta }(\xi ,t).
\label{NPP eq}
\end{equation}%
Here $\mathbf{s}{(\xi ,t)}$ is the position vector of line points labelled
by the arc length ${\xi }$, which varies in interval $0<\xi <2\pi $, the
quantity $\mathbf{\eta }(\xi ,t)$ stands for dissipation, which acts for
marginally small scales. Quantities $\mathbf{s}^{\prime }$ and $\mathbf{s}%
^{\prime \prime }$ are the first and second derivatives of the function $%
\mathbf{s}{(\xi ,t)}$ with respect to variable ${\xi }$. The external
Langevin force $\mathbf{\zeta }(\xi ,t)$ is supposed to be Gaussian with a
correlator concentrated on large scales. An example of evolution of an
initially smooth vortex ring, obeying evolution equation (\ref{NPP eq}), is
depicted in Fig. \ref{loopdeformation}.

One of key quantities in statistical theories is the spectral density tensor
$S_{p}^{\alpha \beta }$, defined as $S_{p}^{\alpha \beta }~\delta
(p+p_{1})~=~\left\langle \mathbf{s}_{p}^{\alpha }\mathbf{s}_{p_{1}}^{\beta
}\right\rangle $ ( $p$ is the one-dimensional wave vector, conjugated to
variable $\xi $). It is responsible for the correlation between tangent
vectors, and for distribution of energy in the space of scale, etc. The
quantity $S_{p}^{\alpha \beta }$ had been calculated in works \cite%
{Nemirovskii1991,Nemirovskii2001b}. The problem has the analytical solution
based on a very popular theoretical trick -- the so called Direct
Interaction Approximation (DIA) for diagramming technique elaborated in
classical turbulence by Wyld \cite{Wyld1961}. According calculations leads
to the result that the spectral density tensor $S_{p}^{\alpha \beta }$ is a
power function of the wave number%
\begin{equation}
S_{p}^{\alpha \beta }~=Cp^{-5},  \label{e:n9}
\end{equation}

Another important result concerns the energy conservation in the $p$ space

\begin{equation}
{\frac{{\partial E_{p}}}{{\partial t}}}~+~{\frac{{\partial P_{p}}}{{\partial
p}}}~=I_{+}(p)~-~I_{-}(p)  \label{e:a1}
\end{equation}%
where $E_{p}={\frac{1}{\sqrt{2\pi }}}\int_{0}^{2\pi }\mathbf{s}_{p}^{\prime }%
\mathbf{s}_{-p}^{\prime }e^{-ip\xi }d\xi $ is the length (energy in LIA) and
$P_{p}^{E}$ is the flux of this quantity in Fourier space (or, equally, in
space of scales). The right-hand side of Eq. (\ref{e:a1}) describes the
creation of additional curvature at a rate of $I_{+}^{K}(p)$ due to external
force and the annihilation of it due to the dissipative mechanism (at a rate
$-I_{-}^{K}(p)$). In a region of wave numbers $p$ remote from both region of
pumping $p_{+}$ and sink $p_{-},\;\;p_{+}~\ll ~p~\ll ~p_{-}$ , the so called
inertial interval, derivative $\partial P_{p}^{E}/\partial p=0$ , so $%
P_{p}^{E}$ is constant equal ,say, $P^{E}.$

Again, the problem stated above, describes one of the main features of
turbulence -- the constant flux of energy in space of scales.

\section{The Kolmogorov type spectra generated by vortex filaments}

Let us discuss how the dynamics of vortex filaments can result in the
Kolmogorov-type spectrum. The energy $E$\ of vortex line configuration $\{%
\mathbf{s}^{\prime }(\xi )\}$ is the integral over all wave vectors $\mathbf{%
k}$ (see \cite{Nemirovskii2002}),%
\begin{equation}
E=\left\langle \frac{{\rho }_{s}{\kappa }^{2}}{2(2\pi )^{3}}\int\limits_{%
\mathbf{k}}\frac{d^{3}\mathbf{k}}{\mathbf{k}^{2}}\int\limits_{0}^{L}\int%
\limits_{0}^{L}\mathbf{s}^{\prime }(\xi _{1})\cdot \mathbf{s}^{\prime }(\xi
_{2})d\xi _{1}d\xi _{2}\exp \left[ i\int\limits_{\xi _{1}}^{\xi _{2}}\mathbf{%
k\cdot s}^{\prime }(\tilde{\xi})d\tilde{\xi}\right] \right\rangle .
\label{E(k) single}
\end{equation}

\begin{figure}[tbp]
\includegraphics[width=11 cm]{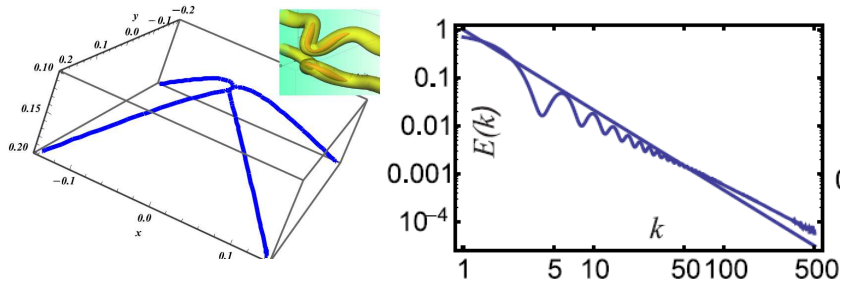}
\caption{(Color online) Left. The touching quasi-hyperbolae describing the
collapsing lines. In the inset we set (as an example) the kinks on the
anti-parallel collapsing vortex tubes obtained in numerical simulation.
Right. The energy spectrum induced by this configuration. }
\label{reconspertum}
\end{figure}

\begin{figure}[tbp]
\includegraphics[width=5 cm]{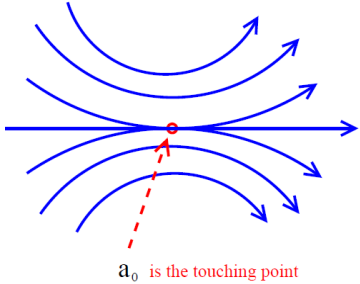}
\caption{(Color online) Schematic picture illustrating the vortex bundle
collapse \protect\cite{Kuznetsov2013}. The initially regular distribution of
vorticity spontaneously concentrates, collapsing in some point $\mathbf{a}%
_{0}$ and forming the singular structure. }
\label{collapse}
\end{figure}

The brackets $<\cdot >$ imply an averaging over the configurations of vortex
loops. I will describe two cases of configurations $\{\mathbf{s}^{\prime
}(\xi )\}$, which indeed lead to the Kolmogorov spectrum. They are the
reconnecting filaments, the collapsing nonuniform vortex bundle, which
appears as a result of vortex dynamics (See. Fig. \ref{reconspertum} and Fig.%
\ref{collapse}). The spectra for these configurations were calculated with
the use of formula (\ref{E(k) single}) in works by the author \cite%
{Nemirovskii2014b,Nemirovskii2015}. From Fig. \ref{reconspertum} it is seen
that for reconnecting lines the spectrum is close to the $E(k)\propto
k^{-5/3}$. The interval of wave numbers where the spectrum $E(k)$ $\approx
k^{-5/3}$ (straight line)is observed, is regulated by the curvature of the
kink and intervortex space $\delta $, which was chosen to be equal to unity.
In reality this spectrum covers a maximum about$\ 1.5$ decades around $%
k\approx $ $2\pi /\delta $. It should be stressed, however, that in the key
numerical works (see references in \cite{Nemirovskii2014b}),\ the ranges for
wave number are also of the order of one decade around $k\approx $ $2\pi
/\delta $.

As for the collapsing nonuniform bundle, then the spectrum $E(k)$ scales as $%
1/k^{1+4/\lambda }$, where $\lambda $ is the increment of nonuniformity of
the vortex lattice on the plane transverse to the bundle (see \cite%
{Nemirovskii2016bundle}). Supposing that the bundle evolves similarly to the
vorticity field in classic hydrodynamics (See e.g., \cite{Kuznetsov2000}),
then $\lambda =6$ and the spectrum $E(k)\propto k^{-5/3}$.

The idea that collapsing singular solutions can play a significant role in
formation of turbulent spectra is being intensively discussed now (see e.g.
\cite{Kuznetsov2000}, \cite{Kerr2013}). The classical examples of such type
spectra are the Phillips spectrum for water-wind waves, created by white
caps -- wedges on water surface.

One more interesting observation is related to the previous paragraph. From
formula \ref{E(k) single} \ it is follows that for the Gaussian loop with
the power-like correlator $\left\langle \mathbf{s}^{\prime }(\xi _{1})\cdot
\mathbf{s}^{\prime }(\xi _{2})\right\rangle \propto (\xi _{1}-\xi
_{2})^{\lambda },$ the energy spectrum is also a power-like function (see
\cite{Nemirovskii2002})
\begin{equation}
E(k)\propto k^{-\frac{2\lambda +2}{\lambda -2}}.  \label{E(k) fractal}
\end{equation}%
Let's apply this consideration to the stochastically deformed vortex loop
described in the previous subsection (See Eq. \ref{e:n9}). The one
dimensional spectrum along the line $\left\langle \mathbf{s}_{p}\mathbf{s}%
_{-p}\right\rangle \propto p^{-5}$ implies that the correlation function $%
\left\langle \mathbf{s}^{\prime }(\xi _{1})\cdot \mathbf{s}^{\prime }(\xi
_{2})\right\rangle $ should scale as $(\xi _{1}-\xi _{2})^{4}$ (See \cite%
{Frisch1995}). Applying this to \cite{Frisch1995} leads to the Kolmogorov
spectrum $E(k)\propto k^{-5/3}$. This result, however, should be taken with
caution. Indeed, the chaotic loop, discussed above is not the Gaussian one,
and knowledge of the correlation function of second order is not enough to
apply relation (\ref{E(k) fractal}) directly.

\section{Free decay of quantum turbulence.}

We submitted several arguments that the main features of classical
turbulence - energy cascade and the Kolmogorov like spectra may appear as a
result of the dynamics of vortex filament. However, one intriguing question
is left. In classical turbulence energy dissipates at the end of inertial
interval due to viscosity. In superfluids viscosity is zero, and the
question arised how the energy, transferred into the region of small scales,
dissapears? Currently, there are several views on the solution to this
paradox.

The first one, initially proposed by Svistunov is connected to a cascade of
nonlinear Kelvin waves. Due to nonlinearity, very small distortions , which
move with very large velocities appear in the system (cf. wih \ref{NPP}). As
it is known from classical hydrodynamics, vortices moving with a velocity $V$
radiate sound, and intensity $I$ of the radiated sound is proportional to
the Mach number in fifth power $I\varpropto (V/c)^{5}$ (c is the sound
velocity). It implies that in order to obtain the significant effect, the
quantized vortex line should have very large curvature or be placed very
near the neighboring vortex or near the boundary.
\begin{figure}[tbp]
\includegraphics[width=14cm]{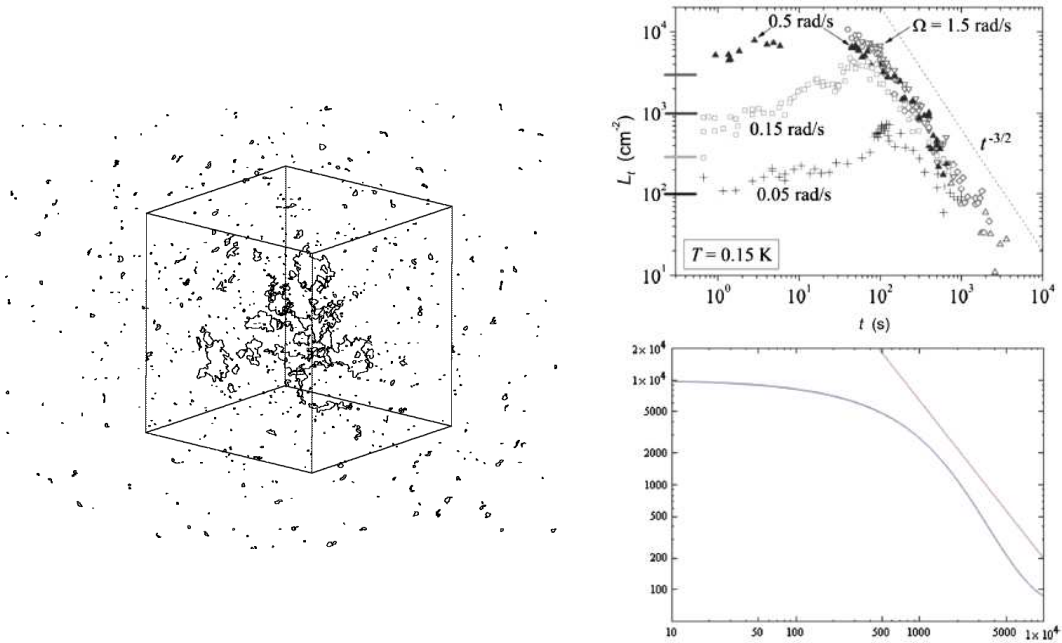}
\caption{(Color online)Left. The emission of small loops from the bulk.
Results of direct numerical simulation (see, \cite{Kondaurova2012}). Right.
Comparison with experiment. In the upper picture there is shown the temporal
attenuation of  the vortex line density, obtained on experiment  \cite{Walmsley2007}.
Temporal behavior of same quantity, calculated from theoretical consideration \cite{Nemirovskii2010}. }
\label{evapcompar}
\end{figure}

Another mechanism is related to the emission of small loops from the bulk.
As discussed earlier, during the reconnection cascade, there appear large
number of very small loops. But these loops, because of their small size
have large mobility and escape from the volume. This process occurs in the
diffusion-like manner, the according calculations agree with experimental
data.

\section{Intermittency}

An interesting phenomena, reminiscent of intermittency, was revealed in work
by Kondaurova et al. \cite{Kondaurova2005}. The vortex tangle does not have
an uniform structure, but, on the contrary consists of various clusters with
high vorticity, which spontaneously appear and disappear in different places
of space.
\begin{figure}[tbp]
\includegraphics[width=14cm]{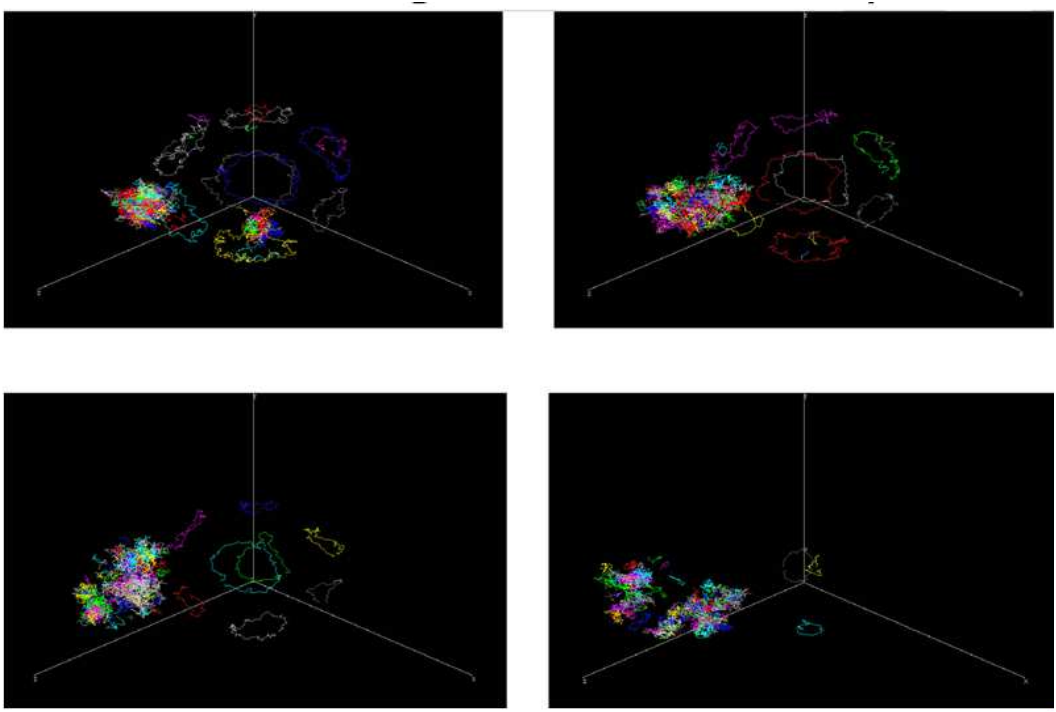}
\caption{(Color online)Schematic picture illustrating the vortex bundle
collapse \protect\cite{Kuznetsov2013}. The regular initially distribution of
vorticity spontaneously concentrates, collapsing in some point $\mathbf{a}%
_{0}$ and forming the singular structure. }
\label{breaking}
\end{figure}

\section{Conclusion and Discussion}

Thus, we demonstrated that some results obtained in quantum fluids, which
were considered as arguments in favour of the idea of modeling classical
turbulence with a set of chaotic quantized vortices, can be explained in the
frame of theoretical models dealing with dynamics of vortex filaments.

The format of the paper, which reflects the presentation at QFS 2016, and
also the page limitation did not allow to cover many other results on this
topic, therefore I restrict myself mainly by my previous results. Among
other theoretical works dealing with dynamics of chaotic vortices I would
like to mention the activity on stochastic nonlinear Kelvin waves and their
role in forming an energy cascade, and dissipation. This topic was
elaborated in works by Vinen \cite{Vinen2000}, Kozik and Svistunov \cite%
{Kozik2009} and in papers by L'vov and Nazarenko with coauthors \cite%
{Lvov2011}. Migdal \cite{Migdal1986} developed the functional formalism for
the study of chaotic vortex filaments with possible application for the
problem of turbulence. The role of chaotic superfluid vortex lines in a
model of turbulent flow was also discussed in the work by Barenghi with
coauthors \cite{Barenghi1997}. In work by Kondaurova \cite{Kondaurova2012}
it was demonstrated how the kinetics of loops resulted in the
quasi-classical decay of quantum turbulence.

This paper should be regarded as an illustration to the idea that the main
properties of turbulence (cascade, scaling laws) are the sequence of the
vortex line dynamics. Now, there appears the following bifurcation. An
optimistic view is that these approaches do relate to the real turbulence
and we have the new vision on processes occurring in turbulent flow. A
pessimistic view is that the phenomena described above have nothing to do
with the real turbulence and all coincidences are occasional.

\bigskip


\end{document}